\documentclass[12pt]{article}
\usepackage{mathrsfs}
\usepackage{amsmath,amssymb,array}
\usepackage{graphicx}

\numberwithin{equation}{section}
\def\p{\partial}

\begin{document}

\begin{titlepage}
\renewcommand{\thefootnote}{\fnsymbol{footnote}}

\begin{center}

\begin{flushright} \end{flushright}
\vspace{2.0cm}

\textbf{\Large{ Constraining the Infrared Behavior of the Soft-Wall \\[0.5cm]
                               AdS/QCD Model }}

\vspace{2cm}

{Ya-Li He}\hspace{0.2cm}and\hspace{0.2cm}{Peng Zhang} \\[0.5cm]


\emph{Institute of Theoretical Physics, College of Applied Sciences, \\
      Beijing University of Technology, Beijing 100124, P.R.China}

\end{center}\vspace{1.5cm}

\centerline{\textbf{Abstract}}\vspace{0.5cm}

By requiring the correct Regge behavior in both meson and nucleon
sectors, we determine the infrared asymptotic behavior of various
background fields in the soft-wall AdS/QCD model, including the dilaton,
the warp factor, and the scalar VEV. We then use a simple
parametrization which smoothly connect these IR limits and their
usual UV limits. The resulting spectrum is compared with
experimental data, and the agreement between them is good.

\end{titlepage}
\setcounter{footnote}{0}


\section{Introduction}

Quantum chromodynamics (QCD) has been established as the genuine theory of strong interaction
for nearly forty years. Quarks and gluons are identified as the fundamental degrees of freedom.
QCD is asymptotically free in the ultraviolet (UV) limit, so people can use standard techniques of
perturbation theory to study the processes with large momentum transfer, like deep inelastic
scatterings, etc. In the infrared (IR) region, however, the coupling constant becomes
strong. Now the effective participants of strong interactions are various hadrons, like $\pi$,
$\rho$, $N$, etc., while quarks and gluons are confined inside these particles. Perturbation theory
cannot be directly used here. People need to develop various effective models to describe the low
energy hadron physics.

AdS/QCD is one of them and has been densely researched for recent several years. This methodology
stems from the idea of the large $N$ expansion due to 't Hooft \cite{tH}, and is directly motivated by
the Anti-de Sitter/conformal field theory (AdS/CFT) correspondence \cite{M, GKP, W} in string theory.
AdS/QCD is a bottom-up approach. It associate QCD operators, like chiral currents and the quark condensate,
to bulk fields propagating in a five-dimensional space, which tends to AdS$_5$ as the fifth coordinate $z$
go to zero. There are mainly two version: the hard-wall model \cite{dTB, EKSS, DP1} and the soft-wall model
\cite{KKSS}. The former can correctly describe the chiral symmetry breaking ($\chi$SB) and low lying hadron states.
The latter is developed for the purpose of realizing the meson Regge behavior due to the linear confinement
in QCD. They find that it is necessary to introduce the dilaton background which is quadratic growth in the
deep IR region $z\rightarrow\infty$.
It is further studied in \cite{GKK} in order to correcly incorporate the $\chi$SB.
AdS/QCD also has interesting relations with the light-front dynamics \cite{BdT1}.
The UV limits of various background fields can be easily fixed. For instance, the warp factor should tend to
that of the AdS space in order to reflect the conformal invariance of the high energy fixed point of QCD,
and the UV behavior of the vacuum expectation value (VEV) of the bulk scalar is determined by the pattern of $\chi$SB.
Therefore works on soft-wall models mainly focus on various improvements in the IR region. However it still seems
arbitrary to some extents.

The main result of this paper is a way to fix the IR asymptotic behavior of various background fields:
the dilaton $\Phi(z)$, the warp factor $a(z)$, and the scalar VEV $v(z)$. We achieve this just by requiring the
model has correct Regge-type spectrum in both meson and nucleon sectors. Nucleons can also be realized \cite{HIY}
in the framework of AdS/QCD by introducing 5D Dirac spinors which correspond to the baryon operators.
In \cite{Z} nucleons are extended to the soft-wall model with asymptotically linear spectrum in both meson and nucleon sectors..
Some other works considering mesons and baryons at the same time can be found in \cite{FBF, VS1, BdT2}.
The main drawback of the model in \cite{Z} is that, although both mesons and nucleons have linear spectra, the spectral
slopes of vectors and that of axial-vectors are different, which is inconsistent with experimental data.
Actually we can argue that it is impossible to improve this if only adjusting the form of the potential and
the scalar VEV. The way we around this is to allow, actually we \emph{have to} allow for being consist with data,
the mass of a bulk field being $z$-dependent.
This idea has also been suggested in the literature, e.g. \cite{CCW, VS2}. Except for conserved currents, a generic operator
will has nonzero anomalous dimension, which is scale-dependent due to the running coupling constant of QCD.
According to the well-known dimension-mass relation, the mass of the corresponding bulk field should be $z$-dependent,
since the fifth coordinate $z$ can be interpreted as the inverse of the 4D energy scale.
What we find is that, by requiring the Regge-type spectrum is properly realized in both meson and nucleon sectors,
the IR asymptotic behaviors of various background fields are totally fixed.
Then using a simple parametrization which smoothly connect these IR limits and their usual UV limits, we can make
predictions and compare them with the observed data. Our philosophy is to reduce the uncertainty of the model as much as
possible by use of known facts.

\section{The model and constraints}

The soft-wall AdS/QCD model is defined in a five-dimensional bulk with the metric
\begin{eqnarray}
ds^2=\,G_{MN}\,dx^M dx^N=a^2(z)\,(\,\eta_{\mu\nu}dx^{\mu}dx^{\nu}-dz^2)\,,\quad  0 < z < \infty\,.
\end{eqnarray}
The factor $a(z)$ is called warp factor, which tends to $z^{-1}$ as $z\rightarrow0$.
There is also a background dilaton $\Phi(z)$ which is assumed to be $O(z^2)$ as $z\rightarrow\infty$,
in order to have Regge-type spectrum in the meson sector \cite{KKSS}.
According to the general rules of the gauge/gravity duality, there are two 5D gauge fields,
$L_M^a$ and $R_M^a$, which correspond to 4D chiral currents $J_{L}^{a\mu}=\bar{q}_L\gamma^{\mu}t^a q_L$ and
$J_{R}^{a\mu}=\bar{q}_R\gamma^{\mu}t^a q_R$. The quark bilinear operator $\bar{q}_L^i q_R^j$ is also an important
4D operator for $\chi$SB. Its holographic dual is a 5D $2\times2$ matrix-valued complex scalar field
$X=(X^{ij})$, which is in the bifundamental representation of the 5D gauge group $SU(N_f)_L \times SU(N_f)_R$ with
$N_f$ being the number of quark flavors. The bulk action for the meson sector is
\begin{eqnarray}
S_M=\int d^4x\,dz\, \sqrt{G}\,e^{-\Phi}\, \mathrm{Tr}\left\{-\frac{1}{4g_5^2}(\,F_L^2+F_R^2)+
    |DX|^2-m_X^2|X|^2\,\right\}\,. \label{SM}
\end{eqnarray}
By matching with QCD, $g_5^2=12\pi^2/N_c=4\pi^2$. The covariant derivative of $X$ is $D_M{X}=\p_M{X}-iL_M{X}+i{X}R_M$.
$F_L$ and $F_R$ are the field strengths of the gauge potentials $L$ and $R$ respectively. The generator $t^a$ is
normalized by $\mathrm{Tr}(\,t^at^b)=\frac{1}{2}\delta^{ab}$.

The bulk scalar $X$ is assumed to have a $z$-dependent VEV: $\langle{X}\rangle=\frac{1}{2}\,v(z)$.
The function $v(z)$ satisfies the equation of motion (EOM)
\begin{eqnarray}
\p_z(\,a^3 e^{-\Phi}\p_z v)-a^5 e^{-\Phi}m_X^2v=0\,. \label{EOMv}
\end{eqnarray}
The mass-square $m_X^2$ may be $z$-dependent due to possible anomalous dimension of $\bar{q}_L q_R$.
From (\ref{EOMv}) we can express  $m_X^2$ as
\begin{eqnarray}
m_X^2=\,\frac{\,v''+(-\Phi'+3a'/a)\,v'}{a^2v}\,\,. \label{msq}
\end{eqnarray}
To describe vector mesons, define $V_M=(L_M+R_M)/2$ and use the axial gauge $V_5=0$.
Expend the field $V_\mu$ in terms of its Kaluza-Klein (KK) modes
$V_\mu(x,z)=\sum_{n}\,\rho_\mu^{(n)}(x)f_V^{(n)}(z)$
with $f_V^{(n)}(z)$ being eigenfunctions of $-\p_5(ae^{-\Phi}\p_5f_V^{(n)})=ae^{-\Phi}M_V^{(n)2}f_V^{(n)}$.
After integrating out the $z$-coordinate, we get an effective 4D action for a tower of massive vector fields
$\rho_\mu^{(n)}(x)$, which can be identified as the fields of $\rho$ mesons with $M_V^{(n)}$ being their masses.
By setting $f_V^{(n)}=e^{\omega/2}\psi_V^{(n)}$,
the equation of eigenfunctions can be transformed to a Schr\"{o}dinger form
$-\psi_V^{(n)\prime\prime}+V_V\psi_V^{(n)}=M_V^{(n)2}\psi_V^{(n)}$ with the potential
\begin{eqnarray}
V_V=\frac{1}{4}\omega'^{\,2}-\frac{1}{2}\omega''\,,\label{VV}
\end{eqnarray}
where $\omega=\Phi-\log{a}$.
Similarly for axial-vectors, define $A_M=(L_M-R_M)/2$. Also use the axial gauge $A_5=0$, expand
$A_\mu(x,z)=\sum_{n}\,a_\mu^{(n)}(x)f_A^{(n)}(z)$, and transform the eigenvalue problem
for $f_A^{(n)}(z)$
into the Schr\"{o}dinger form. The resulting potential $V_A$ for axial-vector mesons is
\begin{eqnarray}
V_A=\frac{1}{4}\omega'^{\,2}-\frac{1}{2}\omega''+\,g_5^2\,a^2v^2\,. \label{VA}
\end{eqnarray}
The corresponding eigenvalue is the mass-square $M_A^{(n)2}$ of the $a_1$ mesons.
Note that there is an additional term $g_5^2\,a^2v^2$, which guarantees the axial-vector resonance
is heavier the vector one with the same radial quantum number.

The spin-1/2 nucleon can also be realized in the AdS/QCD framework  by introducing two 5D Dirac spinors $\Psi_{1,2}$,
which is charged under the gauge fields $L_M$ and $R_M$ respectively. The nucleon sector action is \cite{HIY, Z}
\begin{eqnarray}
S_N&=&\int d^4x\,dz\, \sqrt{G}\,\left(\mathcal{L}_K+\mathcal{L}_I\right)\,, \nonumber\\
\mathcal{L}_K&=&i\overline{\Psi}_1\Gamma^M\nabla_M\Psi_1+i\overline{\Psi}_2\Gamma^M\nabla_M\Psi_2
                -m_{\Psi}\overline{\Psi}_1\Psi_1+m_{\Psi}\overline{\Psi}_2\Psi_2 \,, \label{SN}  \\[0.2cm]
\mathcal{L}_I&=&-g_\mathrm{Y}\overline{\Psi}_1X\Psi_2-g_\mathrm{Y}\overline{\Psi}_2X^\dag\Psi_1\,. \nonumber
\end{eqnarray}
Here $\Gamma^M=e^M_A\Gamma^A=z\delta^M_A\Gamma^A$, and $\{\Gamma^A,\Gamma^B\}=2\eta^{AB}$
with $A=(a,5)$. We choose the representation as $\Gamma^A=(\gamma^a, -i\gamma^5)$ with
$\gamma^5=\mathrm{diag}(I,-I)$.
The covariant derivatives for spinors are $\nabla_M \Psi_1=\p_M\Psi_1+\frac{1}{2}\,\omega^{AB}_M\Sigma_{AB}\Psi_1-iL_M\Psi_1$
and $\nabla_M \Psi_2=\p_M\Psi_2+\frac{1}{2}\,\omega^{AB}_M\Sigma_{AB}\Psi_2-iR_M\Psi_2$.
The only nonzero components of the spin connection
$\omega^{AB}_M$ is $\omega^{a5}_\mu=-\omega^{5a}_\mu=\frac{1}{z}\,\delta^a_\mu$.
The $\mathcal{L}_I$ part introduces the effects of $\chi$SB into the nucleon sector.
In (\ref{SN}) we also allow $m_\Psi$ being $z$-dependent due to possible anomalous dimension of the baryon operator.
Similar with the meson sector, we expand two spinors $\Psi_{a=1,2}$  in terms of its KK modes
\begin{eqnarray}
\Psi_a(x,z)=\begin{pmatrix}\,\, \sum_n N_{L}^{(n)}(x)\,f_{aL}^{(n)}(z)  \,\,\,\, \\[0.2cm]
                           \,\, \sum_n N_{R}^{(n)}(x)\,f_{aR}^{(n)}(z)  \,\,\,\, \end{pmatrix} \,\,.
\end{eqnarray}
The 4D spinors $N^{(n)}=(N_{L}^{(n)}\hspace{-0.1cm}, N_{R}^{(n)})^{\mathrm{T}}$ are interpreted as nucleon fields.
The internal wave functions $f$'s satisfy four coupled 1st order differential equations.
By acting one more derivative and eliminating two right-handed $f$'s, we get a coupled Sterm-Liouville
eigenvalue problem for $f_{L}^{(n)}\equiv(f_{1L}^{(n)},f_{2L}^{(n)})^\mathrm{T}$.
Define $\chi_{L}^{(n)}=a^{2}f_{L}^{(n)}$, the coupled Schr\"{o}dinger equation for $\chi_{L}^{(n)}$ is
$-\chi_L^{(n)}{''}+V_N\chi^{(n)}_L=M_N^{(n)2}\chi^{(n)}_L$. The potential matrix $V_N$ is
\begin{eqnarray}
V_N
    =\begin{pmatrix}\,\, m_{\Psi}^2a^2+(m_{\Psi}a)'+u^2 & u' \\[0.2cm]
    u' & m_{\Psi}^2a^2-(m_{\Psi}a)'+u^2 \,\,\,\end{pmatrix}\,,\label{VN}
\end{eqnarray}
with $u(z)=\frac{1}{2}\,g_\mathrm{Y}av$. The eigenvalue $M_N^{(n)2}$ is the mass-square of nucleon and its radial excitations.

Now we start to analyze the asymptotic behavior of various background fields in the model, i.e.
the dilaton $\Phi(z)$, the warp factor $a(z)$, and the scalar VEV $v(z)$. The UV limit is relatively simple
to argue. For the warp factor
\begin{eqnarray}
a(z)\sim \,\frac{L}{z}\,,\qquad z\rightarrow0\,. \label{aUV}
\end{eqnarray}
This is because of the conformal invariance of the UV fixed point. So the bulk space should be asymptotic 5D AdS.
The value of the characteristic length $L$ does not affect the resulting spectrum. 
For the scalar VEV
\begin{eqnarray}
v(z)\sim \,Az+Bz^3\,,\qquad z\rightarrow0\,. \label{vUV}
\end{eqnarray}
The linear term corresponds to the explicit $\chi$SB due to the quark mass, while the cubic term describes the
spontaneous breaking by the nonzero quark condensate. Unlike the warp factor and the scalar VEV,
the UV limit of the dilaton, however, cannot be uniquely fixed. The reason is as follows. Since QCD is
asymptotically free, the conformal dimension of any operator, at the UV fixed point, is just its classical value,
which is 3 for $\bar{q}_L q_R$. Therefore by the mass-dimension relation $m_X^2=\Delta(\Delta-4)$, we have
\begin{eqnarray}
m_X^2(z)\sim\,-3\,,\qquad z\rightarrow0\,. \label{mXUV}
\end{eqnarray}
From the expression (\ref{msq}) we can see that the above equation (\ref{mXUV}) holds if and only if
$\Phi(z)\sim z^\alpha$ as $z\rightarrow0$ with $\alpha>0$. Actually the UV limit of the dilaton could also
depend on the form of the scalar potential in the bulk action \cite{GKK, Z}.  So it is, generally speaking,
model-dependent.

Having studied the UV behavior, we now turn to the IR. For the dilaton it must be
\begin{eqnarray}
\Phi(z)\sim\,O(z^2)\,,\qquad z\rightarrow\infty\,,
\end{eqnarray}
which guarantees $M_V^{(n)2}\hspace{-0.1cm}\sim O(n)$ as $n\rightarrow\infty$ for vector mesons \cite{KKSS}.
Suppose $a(z)\sim O(z^{\gamma})$ as $z\rightarrow\infty$,\footnote{If assume non-power function, e.g. $a\sim e^z$, it will
destroy the linear spectrum in the nucleon sector.} we always have $(\log{a}\hspace{-0.05cm})\,'\hspace{-0.1cm}\sim O(z^{-1})$
for any power $\gamma$. Therefore, from the expression (\ref{VV}) of the potential, only considering vector mesons cannot give any constraint
on the IR behavior of the warp factor $a(z)$. One of key observations of this paper is that we can fixed that by the nucleon sector.
Look at the nucleon potential matrix (\ref{VN}), please. At the IR fixed point, QCD becomes a strongly coupled, but well-defined,
conformal field theory. So the dimension of the baryon operator should be finite, which means that $m_\Psi$, although may be
$z$-dependent, must tend to a finite constant as $z\rightarrow\infty$. Consider the diagonal terms of (\ref{VN}) first.
$m_{\Psi}^2a^2$ must dominates $\pm(m_{\Psi}a)'$ when $z$ large. However which one is dominant between the 1st term $m_{\Psi}^2a^2$
and the 3rd term $u^2$ is a crucial issue. We determine this by reduction to the absurd. Suppose $u^2\propto{a^2v^2}$ dominates,
then the asymptotic linearity of nucleon spectrum forces $a^2v^2\sim O(z^2)$ in the IR. However note that $u^2\propto{a^2v^2}$
also appears in the axial potential (\ref{VA}), there will be another $O(z^2)$ term in addition to the 1st term  in $V_A(z)$.
Then this implies the axial-vector spectrum, although still linear, has a different slope with vector mesons, which is inconsistent
with experimental data. Therefore the conclusion is: $m_{\Psi}^2a^2$ dominates $u^2$. Again by the spectral linearity of nucleons,
we obtain the desired IR limit of the warp factor as
\begin{eqnarray}
a(z)\sim\,O(z)\,,\qquad z\rightarrow\infty\,. \label{aIR}
\end{eqnarray}
By requiring vectors and axial-vectors have the same spectral slopes, we only know $v(z)\sim O(z^{1-\varepsilon})$ at IR for
some positive $\varepsilon$. To further constrain it, we suppose the chiral symmetry is not asymptotically restored
\cite{SV}.\footnote{There are some controversies about this issue among experts, see e.g. \cite{Gl} for the opposite opinion.}
This means that $V_A-V_V\propto a^2v^2$ should tend to some nonzero constant as $z\rightarrow\infty$. Therefore the
IR behavior of the scalar VEV is
\begin{eqnarray}
v(z)\sim\,O(z^{-1})\,,\qquad z\rightarrow\infty\,. \label{vIR}
\end{eqnarray}
As a cross check, we find $u'\propto(av)'$ tends to zero in the deep IR region. Two Schr\"{o}dinger equations for nucleons
decouple with each other, which means the mass-difference between a nucleon state and its parity partner becomes smaller
and smaller.\footnote{This asymptotic degeneracy of nucleon states in a parity doublet does not necessarily imply,
at least theoretically, the chiral symmetry restoration, see e.g. \cite{JPS, K}.} This is consistent with the observed data.

Additionally it can be shown that, with these IR limits, the scalar and pseudoscalar mesons also have asymptotically
linear spectral trajectories  parallel to those of vectors and axial-vectors.
By directly applying the method of \cite{KKSS}, it can be further shown that the relation between the mass-square and
the total angular-momentum quantum number $J$ for higher spin mesons is indeed Regge-type.
These facts exhibit the consistency of the AdS/QCD model and our asymptotic relations (\ref{aIR}) and (\ref{vIR})
which are our main results in the present work.

\section{A simple parametrization}

Having determined various asymptotic behaviors of background fields, we will use simple parametrizations which smoothly
connect these asymptotes from UV to IR. First we simply choose
\begin{eqnarray}
\Phi=\kappa^2z^2\,. \label{Phi}
\end{eqnarray}
It is shown in \cite{KKSS2} that the sign of the dilaton should be positive to avoid a spurious
massless state in the vector sector.
Since we allow $m_X^2$ to be $z$-dependent, the choice (\ref{Phi}) does not raise difficulties for the correct
realization of $\chi$SB \cite{VS2}. We parametrize the warp factor and the scalar VEV as
\begin{eqnarray}
a(z)&=&\,\frac{\,1+\mu z^2}{z}\,\,,\\[0.1cm] \label{a}
v(z)&=&\,\frac{\,Az+Bz^3}{1+Cz^4}\,\,.  \label{v}
\end{eqnarray}
We have chosen the characteristic length $L$ of AdS$_5$ to be 1.
All of these three parametrizations have correct UV and IR behaviors determined in the previous section.
By use of (\ref{msq}) the $z$-dependence of $m_X^2$ has been fixed. Since the dilaton $\Phi$ has positive power,
$m_X^2$ has correct UV limit, i.e. $m_X^2\sim-3$. In the deep IR it can be shown that $m_X^2$ tends to zero.
With these parametrizations we can numerically solve the Schr\"{o}dinger equation with the potentials $V_V(z)$
in (\ref{VV}) and $V_A(z)$ in (\ref{VA}). By fitting the vector meson masses and those of the axial-vectors,
we choose the values of the five parameters as follows
\begin{eqnarray}
&\kappa=415.9\,\mathrm{MeV}\,,\quad \mu=860.4\,\mathrm{MeV}\,;& \nonumber \\
&A=2.1\,\mathrm{MeV}\,,\quad B=(411.9\,\mathrm{MeV})^3\,,\quad C=(733.6\,\mathrm{MeV})^4\,.& \label{para_m}
\end{eqnarray}
Vector mesons and axial-vector mesons both have asymptotically linear mass-squares,
with the same slope $4\kappa^2$.
The resulting spectra together with the observed values are listed in Table \ref{rho} and \ref{a1} respectively.
The agreement between them is good, especially for higher resonance states.

\begin{table}
\centering
\begin{tabular}{|c|c|c|c|c|c|c|c|c|}
\hline
$\rho$                   &  0  &  1   &  2   &  3   &  4   &  5   &  6   \\
\hline
$m_{\mathrm{th}}$ (MeV)  &1003 & 1306 & 1550 & 1759 & 1947 & 2118 & 2276 \\
\hline
$m_{\mathrm{ex}}$ (MeV)  &775.5& 1465 & 1570 & 1720 & 1909 & 2149 & 2265 \\
\hline
error                   &29.3\%& 10.8\% &1.3\% &2.3\%&2.0\%& 1.4\%&0.5\% \\
\hline
\end{tabular}
\caption{\small{The theoretical and experimental values of vector meson masses.}}\label{rho}
\end{table}

\begin{table}
\centering
\begin{tabular}{|c|c|c|c|c|c|c|}
\hline
$a_1$                    &  0  &  1   &  2   &  3   &  4   &  5    \\
\hline
$m_{\mathrm{th}}$ (MeV)  &1452 & 1646 & 1842 & 2022 & 2187 & 2340  \\
\hline
$m_{\mathrm{ex}}$ (MeV)  &1230 & 1647 & 1930 & 2096 & 2270 & 2340  \\
\hline
error                   &18.1\%& 0.0\% &4.6\% &3.6\%& 3.7\%& 0.0\% \\
\hline
\end{tabular}
\caption{\small{The theoretical and experimental values of axial-vector meson masses.}}\label{a1}
\end{table}

For nucleons we parametrize the bulk spinor mass $m_\Psi$ as
\begin{eqnarray}
m_\Psi=\,\frac{\,\frac{5}{2}+\mu_1z\,}{\,1+\mu_2z\,}\,. \label{mpsi}
\end{eqnarray}
This parametrization gives the correct UV limit $5/2$ which correspond to the classical dimension $9/2$
of the baryon operator by the mass-dimension relation for spinors $m_\Psi=\Delta-2$. At IR $m_\Psi$ tends to
a constant $\mu_1/\mu_2$ which, together with the parameter $\mu$ in the warp factor, determines the mass-square slope for nucleons.
It is needed to numerically solve the coupled Schr\"{o}dinger equation with proper boundary conditions \cite{Z}.
We simply fix $\mu_1=1$GeV, while $\mu_2$ and the Yukawa coupling constant $g_\mathrm{Y}$ are chosen as
\begin{eqnarray}
g_\mathrm{Y}=9.2\,,\quad \mu_2=4573\,\mathrm{MeV}\,.
\end{eqnarray}
The calculated nucleon masses and the corresponding data are listed in Table \ref{nucl}.
The agreement between them is also reasonable.
\begin{table}
\centering
\begin{tabular}{|c|c|c|c|c|c|c|c|}
\hline
$N$                     &  0     &   1    &  2     &  3     &  4     &  5     &   6  \\
\hline
$m_{\mathrm{th}}$(MeV)  & 937   & 1434   & 1583   & 1783  & 1842   & 2029   & 2065 \\
\hline
$m_{\mathrm{ex}}$(MeV) & 939    & 1440   & 1535   & 1650   & 1710   & 2090   & 2100 \\
\hline
error                   &0.2\%   &0.5\%   &3.2\%   &8.0\%   &7.7\%   &2.9\%   &1.7\% \\
\hline
\end{tabular}
\caption{\small{The theoretical and experimental values of the spin-1/2 nucleon masses.}}\label{nucl}
\end{table}

\section{Conclusions}

By requiring the model has correct Regge-type spectrum in both meson and nucleon sectors,
we determine the IR behavior of various background fields in the soft-wall AdS/QCD model,
including the dilaton $\Phi(z)$, the warp factor $a(z)$, and the scalar VEV $v(z)$.
More precisely, our arguments are mainly based on: (i) $M_n^2\sim O(n)$ as $n\rightarrow\infty$
for both mesons and nucleons. (ii) The meson spectral slopes are asymptotically equal.
(iii) The distance between the mass-squares of a vector resonance and the corresponding axial-vector
resonance tends to a finite nonzero constant. We use simple parametrizations which smoothly connect
these IR limits with their usual UV limits. The agreement between the predictive value and experimental
data is good. In the present work we restrict ourself to the lowest order effective bulk action.
Whether our arguments could supply some constraints on higher dimensional terms, e.g. scalar potentials,
is an interesting further issue. There is an other way \cite{PW}, which is more economic theoretically,
to realize baryon states in the framework of holographic QCD by a five-dimensional skyrmion. It is 
interesting to see how to realize the linear spectra for baryons in this model and which constraints   
imposed on various background fields.

\end{document}